\documentclass[prl,twocolumn,showpacs]{revtex4}

\pdfoutput=1

\usepackage{amsmath,amssymb}
\usepackage{graphicx}
\usepackage{dcolumn}

\newcommand{\hxi}{\hat{\xi}}
\newcommand{\bJ}{\overline{J}}
\newcommand{\fpi}{F}

\begin{document}

\title{Individual complex Dirac eigenvalue distributions from random
  matrix theory and comparison to quenched lattice QCD with a quark
  chemical potential}

\author{G. Akemann,$^1$ J. Bloch,$^2$ L. Shifrin,$^1$ and T.
  Wettig$^2$}

\affiliation{$^1$Department of Mathematical Sciences \& BURSt Research
  Centre, Brunel University West London, Uxbridge UB8 3PH, United
  Kingdom\\ 
  $^2$Institute for Theoretical Physics, University of Regensburg,
  93040 Regensburg, Germany}

\date{November 21, 2007}

\begin{abstract}
  We analyze how individual eigenvalues of the QCD Dirac operator at
  nonzero quark chemical potential are distributed in the complex
  plane.  Exact and approximate analytical results for both quenched
  and unquenched distributions are derived from non-Hermitian random
  matrix theory. When comparing these to quenched lattice QCD spectra
  close to the origin, excellent agreement is found for zero and
  nonzero topology at several values of the quark chemical potential.
  Our analytical results are also applicable to other physical systems
  in the same symmetry class.
\end{abstract}



\maketitle

\textbf{Introduction.}  Hermitian random matrix theory (RMT), which
describes systems with real spectra, enjoys many applications in
physics and beyond.  Dropping the Hermiticity constraint results in
matrices whose eigenvalues are, in general, complex.  Examples are the
Ginibre ensembles \cite{Ginibre} or their chiral counterparts
\cite{Halasz:1997fc}.  Although these ensembles describe non-Hermitian
operators, they have found many applications (see \cite{FS} for a
recent review), ranging from dissipation in quantum maps \cite{GHS88}
over quantum chromodynamics (QCD) at nonzero quark chemical potential
\cite{Stephanov:1996ki} to the brain auditory response described by
nonsymmetric correlation matrices \cite{KDI}.

Observables that are typically computed in RMT are spectral
correlation functions.  Alternatively, one can study the distributions
of individual eigenvalues, provided that the latter can be ordered.
For RMT with real eigenvalues, all such distributions are known and
have found a variety of important applications. For example, the
largest eigenvalue follows the Tracy-Widom distribution
\cite{Tracy:1992rf} and appears in the longest increasing sub-sequence
of partitions \cite{BDJ} or growth processes \cite{PS}.  The smallest
eigenvalue distribution in chiral RMT has become a standard tool in
lattice QCD to extract the low-energy constant (LEC) $\Sigma$ that
appears in chiral perturbation theory (chPT) and is related to the
chiral condensate \cite{Fukaya:2007fb}.  This distribution is also
sensitive to the gauge field topology and can be used to distinguish
different patterns of chiral symmetry breaking \cite{Edwards:1999ra}.

In this paper, we generalize some of these results to the case of
non-Hermitian chiral RMT in the unitary symmetry class.  We study the
distributions of individual eigenvalues in the complex plane and
derive analytical results for the chiral RMT introduced in
Ref.~\cite{Osborn:2004rf}.  Our main focus will be on QCD, but our
findings are also relevant for other systems with complex eigenvalues
in the same symmetry class.

In QCD, a nonzero quark chemical potential $\mu$ leads to a complex
spectrum of the Dirac operator. In the large-volume limit, chiral RMT
is equivalent \cite{Basile:2007ki} to the chiral effective theory for
the epsilon-regime of QCD \cite{Gasser:1987ah}, which is a particular
low-energy limit of the full theory. Here, a virtue of $\mu\neq0$ is
that $\mu$ couples to the second LEC in leading order of chPT, $\fpi$,
which is related to the pion decay constant \cite{Toublan:1999hx}.  A
comparison of lattice QCD data to individual complex Dirac eigenvalue
distributions from RMT thus allows us to determine both $\Sigma$ and
$\fpi$ (for related methods, see
Refs.~\cite{Damgaard:2005ys,Osborn:2005zi}).

Unfortunately, lattice QCD with dynamical fermions at $\mu\ne0$ faces
a serious difficulty due to the loss of reality of the action.  It is
very hard to obtain significant statistics in unquenched simulations,
and therefore we will only compare to quenched simulations below.
However, for $\mu<m_\pi/2$ or $\mu^2F^2V<1$ (where $m_\pi$ is the pion
mass and $V$ is the volume) the sign problem is not severe
\cite{Splittorff:2006fu:2007ck}, and our method can be used to
determine $F$ from such unquenched lattice data.  Therefore we also
derive RMT results for unquenched QCD, thus adding to the predictions
for spectral densities \cite{Akemann:2004dr} and the average phase
factor \cite{Splittorff:2006fu:2007ck}.

What is known from RMT for individual eigenvalue correlations in the
complex plane?  For the non-chiral, unitary Ginibre ensemble the
repulsion (or spacing distribution) of complex levels was computed in
\cite{GHS88} and successfully compared to lattice QCD data in the bulk
of the spectrum \cite{Markum:1999yr}.  For maximal non-Hermiticity,
the distribution of the largest eigenvalue with respect to radial
ordering is also known \cite{EK}.  However, in QCD it is the
eigenvalues closest to the origin that carry information about
topology and LECs, and therefore we concentrate on these in the
following.

The complex spectral correlation functions of the QCD Dirac operator
at $\mu\ne0$ were computed from different (but equivalent) chiral RMTs
in Refs.~\cite{Splittorff:2003cu,Osborn:2004rf,Akemann:2004dr} and
compared to quenched lattice QCD in
Refs.~\cite{Akemann:2003wg,Osborn:2005zi}. Later, a Dirac operator
with exact chiral symmetry at $\mu\neq0$ was constructed
\cite{Bloch:2006cd,Bloch:2007xi} and tested against chiral RMT for
topological charge $\nu=0,1$.  Here, we compare the data of
Ref.~\cite{Bloch:2006cd} to our newly derived individual complex
eigenvalue distributions, resulting in a much improved signal.  For a
recent review
we refer to Ref.~\cite{Akemann:2007rf}.\\[-3mm]

\textbf{Complex eigenvalue distributions.}  We start by defining the
gap probability and the distribution of an individual eigenvalue in
the complex plane.  Suppose the partition function $\mathcal Z$ can be
written in terms of $N$ complex eigenvalues $z_j$ of some operator,
with a joint probability distribution function (jpdf) $ {\mathcal
  P}(\{z\})$, symmetric in all its arguments, to be specified.  (For
simplicity, we consider only jpdf's with additional symmetry
$z\leftrightarrow -z$, restricting ourselves to the upper half-plane
$\mathbb{C}_+$.)  The complex eigenvalue density correlation functions
are defined as
\begin{equation}
  R_k(z_1,\ldots,z_k) \equiv \frac{1}{{\mathcal Z}} \frac{N!}{(N-k)!} 
  \prod_{j=k+1}^N\int_{\mathbb{C}_+} \!\! d^2z_j {\mathcal P}(\{z\})\:.
  \label{Rkdef}
\end{equation}
The simplest example $R_1(z)$ is just the spectral density.  The gap
probability $E_k[J]$ is defined as the probability that there are
exactly $k$ eigenvalues inside the set $J$ and $N-k$ eigenvalues in
its complement $\bJ\equiv\mathbb{C}_+\slash J$,
\begin{equation}
  E_k[J]\equiv\frac{1}{{\mathcal Z}}\frac{N!}{(N-k)!}
  \prod_{j=1}^k \int_{J}\!d^2z_j\!\!
  \prod_{i=k+1}^N\int_{\bJ}\!d^2z_i\,{\mathcal P}(\{z\})\:.
  \label{Edef}
\end{equation}
If all $R_k$ are known, the gap probabilities follow as in the real
case \cite{Akemann:2003tv},
\begin{equation}
  E_k[J]=\sum_{\ell=0}^{N-k}\frac{(-1)^\ell}{\ell!}\prod_{j=1}^{k+\ell}
  \int_{J}d^2z_j\,R_{k+\ell}(z_1,\ldots,z_{k+\ell})\:.
  \label{Eexp}
\end{equation}
Parameterizing the boundary $\partial J$ of $J$ in ${\mathbb{C}_+}$ as
$z(\tau)=x(\tau)+i\,y(\tau)$, we can define the probability
$p_k(J,\tau)$ for \mbox{$k-1$} eigenvalues to be inside $J$, for the
eigenvalue $z_k=z(\tau)$ to be on the contour $\partial J$ at $\tau$,
and for $N-k$ eigenvalues to be in the complement $\bJ$,
\begin{equation}
  \label{pdef}
  p_k(J,\tau)\equiv \frac{k}{\mathcal Z} 
  \binom {N} {k} \prod_{j=1}^{k-1}\int_{J} \!d^2z_j\!\!\!
  \prod_{i=k+1}^N\int_{\bJ} \!d^2z_i{\mathcal P}(\{z\})\big|_{z_k=z(\tau)}\:.
\end{equation}
(Because eigenvalues repel each other in RMT, the probability of
finding two eigenvalues at $z(\tau)\neq0$ is zero.)  An ordering on
$\mathbb{C}_+$ is induced by a family of sets of increasing area with
mutually nonintersecting contours.  Via the Riemann mapping theorem,
this can always be reduced to radial ordering.  Definitions
\eqref{Edef} and \eqref{pdef} are related through a variational
derivative,
\begin{equation}
  \frac{\delta E_k[J]}{\delta z(\tau)} = 
  k!\big[p_k(J,\tau)-p_{k+1}(J,\tau)\big]\:.
  \label{Eprel}
\end{equation}
Employing the expansion \eqref{Eexp}, we can express the $p_k(J,\tau)$
through densities.  For example, for the first eigenvalue,
\begin{align}
  p_1&(J,\tau) = 
  R_1(z(\tau))-\int_J d^2z_1 \, R_2(z_1,z(\tau))\notag\\
  &+\frac{(-1)^2}{2!}\int_Jd^2z_1\int_Jd^2z_2\, R_3(z_1,z_2,z(\tau))+\ldots
  \label{p1exp}
\end{align}

\textbf{Results from RMT.}  The above considerations hold for any
jpdf, including the jpdf appearing in the lattice QCD partition
function in terms of complex Dirac operator eigenvalues and the jpdf
of chiral RMT.  We now consider the latter.  The RMT for unquenched
QCD with $\mu\ne0$ \cite{Osborn:2004rf} we use here is given by
\begin{equation}
  {\mathcal P}(\{z_i\})= \prod_{j=1}^N w^{(N_f,\nu)}(z_j)
  |\Delta_N(\{z^2\})|^2\:.
  \label{Prmt}
\end{equation}
The Vandermonde, $\Delta_N(\{z^2\})= \prod_{i>j}^N(z_i^2-z_j^2)$,
coming from the diagonalization of complex matrices of dimension
$N\times(N+\nu)$ (we take $\nu\ge0$ for convenience), leads to a
repulsion of eigenvalues.  (For the chiral RMTs corresponding to
adjoint or two-color QCD, the Jacobians will be different, leading to
different patterns of eigenvalue repulsion, see, e.g.,
Ref.~\cite{Akemann:2007rf}.)  The weight $w$ depends on $N_f$
dynamical quark flavors with masses $m_f$ ($f=1,\ldots,N_f$) and on
the number $\nu$ of exactly zero eigenvalues (corresponding to the
topological charge),
\begin{align}
  &w^{(N_f,\nu)}(z_j)=
  \prod_{f=1}^{N_f} (m_f^2-z_j^2)
  \label{wdef}\\
  &\quad\times|z_j|^{2\nu+2}
  K_\nu\!\left(\frac{N(1+{\hat\mu}^2)}{2{\hat\mu}^2}|z_j|^2\right)
  e^{\frac{N({\hat\mu}^2-1)}{4{\hat\mu}^2}\left(z_j^2+z_j^{*\,2}\right)}\:,
  \notag
\end{align}
where $K_\nu$ is a modified Bessel function and $\hat\mu$ is the
chemical potential in the random matrix model.  The first factor in
Eq.~\eqref{wdef} originates from the Dirac determinants. The
non-Gaussian weight function results from an integration over angular
and auxiliary variables \cite{Osborn:2004rf}.  For $\hat\mu\to 0$ the
$z_k$ are back on the imaginary axis.  Complex RMT yields the
following result for the densities \cite{Akemann:2002ym:2002js},
\begin{equation}
  R_k(z_1,\ldots,z_k)
  =\prod_{\ell=1}^k w^{(N_f,\nu)}(z_\ell)
  \det_{1 \leq i,j \leq k}K_{N}(z_i,z^*_j)\:,
  \label{Rksol}
\end{equation}
given in terms of the kernel $K_N(z_i,z_j^*)$ of (bi-)orthogonal
polynomials with respect to the weight of Eq.~\eqref{wdef}.  In the
quenched case ($N_f=0$), these are given by Laguerre polynomials in
the complex plane \cite{Osborn:2004rf}.  All unquenched density
correlations are given explicitly in Ref.~\cite{Akemann:2004dr}.  A
determinental expression follows for the $E_k[J]$ in terms of the
kernel operator times the characteristic function of $J$.
Eq.~\eqref{Eexp} is called its Fredholm determinant expansion.

As mentioned above, in the limit of large volume $V$, RMT is
equivalent to QCD in the epsilon-regime \cite{Basile:2007ki}.  In this
regime, the chemical potential, the quark masses, and the Dirac
eigenvalues are rescaled such that the parameters $\alpha\equiv
2N{\hat\mu}^2\,(=V\fpi^2\mu^2)$, $\eta_f\equiv Nm_f\,(=V\Sigma m_f)$,
and $\xi_k\equiv Nz_k\,(=V\Sigma z_k)$ stay finite in the large-$N$
(large-$V$) limit.  In parentheses, we have given the scaling of these
parameters in terms of the LECs of chPT.\\[-3mm]

\begin{figure}[t]
  \centering
  \includegraphics[width=43mm]{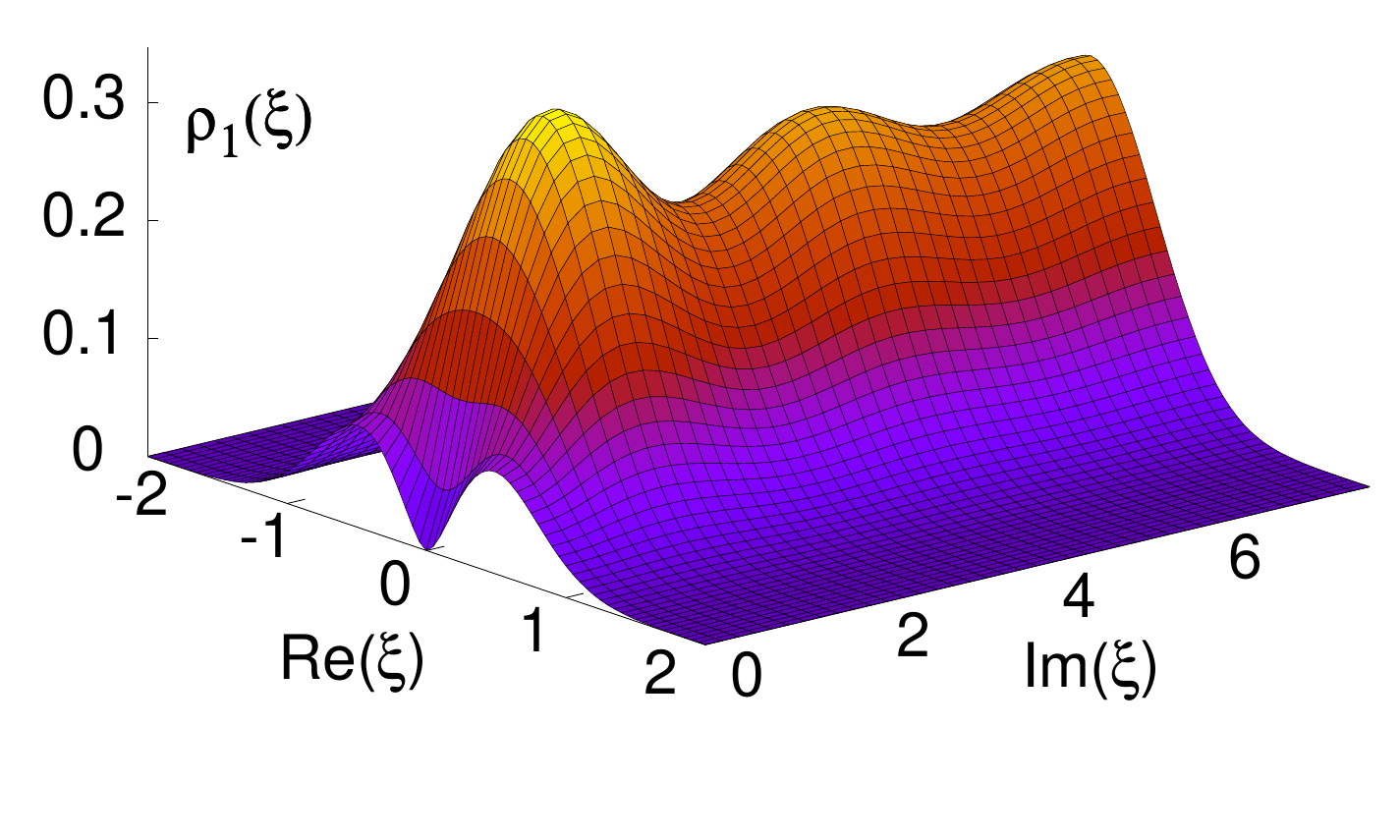}\hfill%
  \includegraphics[width=43mm]{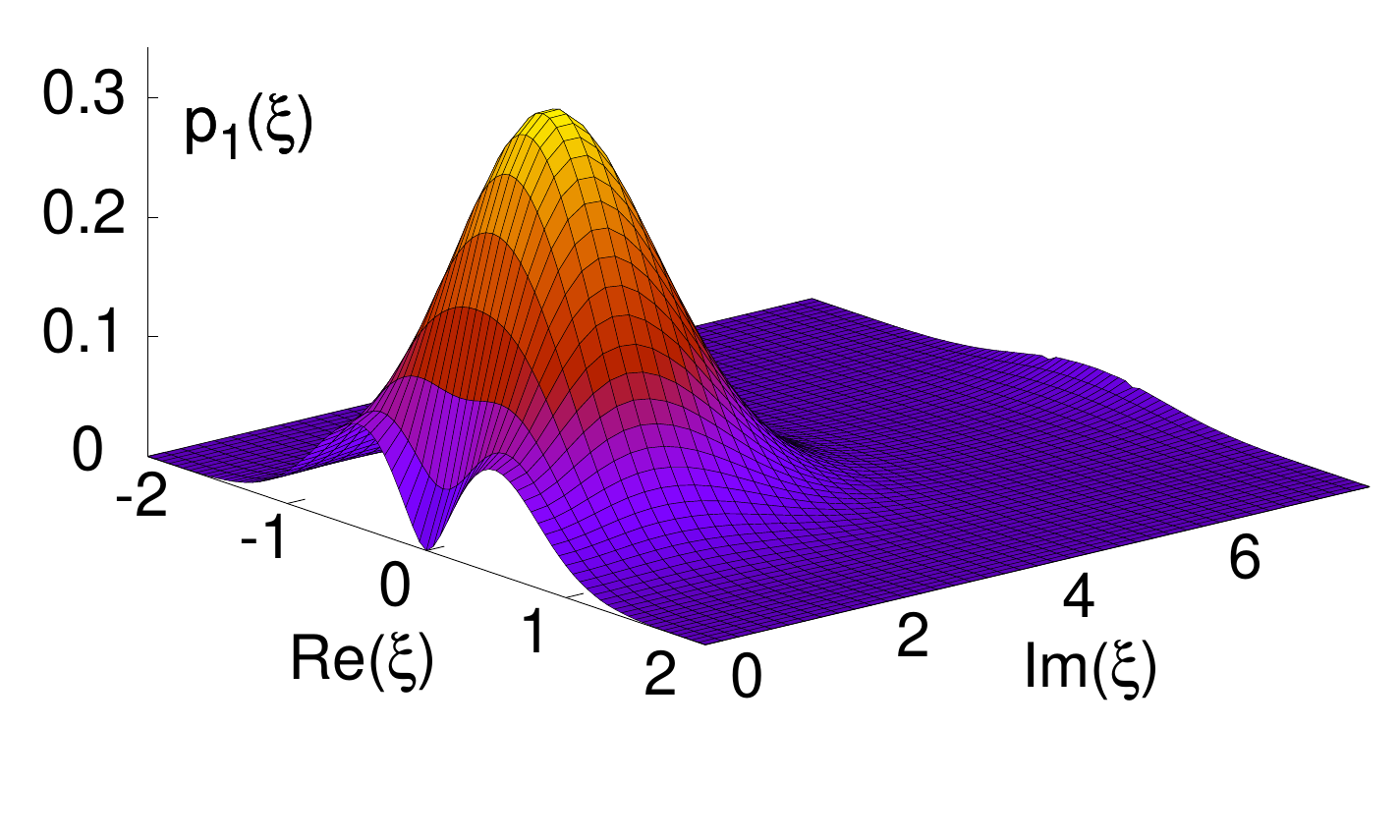}
  \vspace*{-7mm}
  \caption{Quenched density $\rho_1(\xi)$ of Eq.~\eqref{rhoQ} (left),
    and quenched $p_1(\xi)$ from Eq.~\eqref{p1exp} (including the
    first three terms) for circular $J$ (right), both for $\nu=0$ and
    $\alpha=0.174$.}
  \label{3d}
\end{figure}

\textbf{Quenched case.} In the quenched case, the RMT result for the
microscopic spectral density
$\rho_1(\xi)\equiv\lim_{N\to\infty}R_1(\xi=z/N)/N$ is given by
\cite{Splittorff:2003cu,Osborn:2004rf}
\begin{equation}
  \rho_1(\xi)= 
  \frac{|\xi|^2K_\nu\left(\frac{|\xi|^2}{4\alpha}\right)}{2\pi\alpha}
   e^{\frac{-\xi^2-\xi^{*\,2}}{8\alpha}}      
  \!\int_0^1\!\!dt\,t\, e^{-2\alpha t^2}|I_\nu(t\xi)|^2,
  \label{rhoQ}
\end{equation}
where $I_\nu$ is a modified Bessel function.  The rescaled kernel
giving all correlation functions according to Eq.~\eqref{Rksol} was
derived in Refs.~\cite{Osborn:2004rf,Akemann:2004dr}.  In
Fig.~\ref{3d} we show as an example the density $\rho_1(\xi)$ and the
distribution $p_1(\xi)$ of the first eigenvalue from Eq.~\eqref{p1exp}
(in which $J$ is chosen to be semi-circular and only the first three
terms are included).  As in the case of real eigenvalues
\cite{Akemann:2003tv}, we see that the expansion converges rapidly.
Higher-order terms merely assure that $p_1(\xi)$ remains zero for
large $|\xi|$.

For increasing $\alpha$, the quenched density Eq.~\eqref{rhoQ} rapidly
becomes rotationally invariant close to the origin.  In terms of the
new variable $\hat{\xi}=\xi/2\sqrt{\alpha}$, it becomes
\begin{equation}
  \rho_1(\hxi)\overset{\alpha\to\infty}=
  \frac{2|\hxi|^2}{\pi} K_\nu(|\hxi|^2)I_\nu(|\hxi|^2)\:.
  \label{rhoQS}
\end{equation}
In this limit, we can derive a closed expression for the gap
probability \cite{ASII}.  Because of the rotational symmetry we choose
$J$ to be a semi-circle of radius $r\equiv|\hat\xi|$ and obtain
\begin{align}
  E_0(r)&=\prod_{\ell=0}^{\infty}\biggl\{ 
  \frac{r^{4\ell+2\nu+2} K_{\nu+1}(r^2)}{2^{2\ell+\nu}\ell!(\ell+\nu)!}
  \label{E0}\\
  &+ r^2\left[ K_{\nu+1}(r^2)I_{\nu+2}^{[\ell-2]}(r^2)    
  +    K_{\nu+2}(r^2)I_{\nu+1}^{[\ell-1]}(r^2) \right]
  \biggr\}\:,\notag
\end{align}
where we have introduced the incomplete Bessel function
$I_{\nu}^{[\ell]}(x)\equiv\sum_{n=0}^{\ell}(x/2)^{2n+\nu}/n!(n+\nu)!$
for $\ell\ge 0$, and zero otherwise.  Our quenched expression
Eq.~\eqref{E0} generalizes the corresponding result of
Ref.~\cite{GHS88} for the non-chiral Ginibre ensemble, which is given
in terms of incomplete exponentials $e_\ell(x)=\sum_{n=0}^\ell
x^n/n!$.

\begin{figure}[t]
  \centering
  \includegraphics[width=42mm]{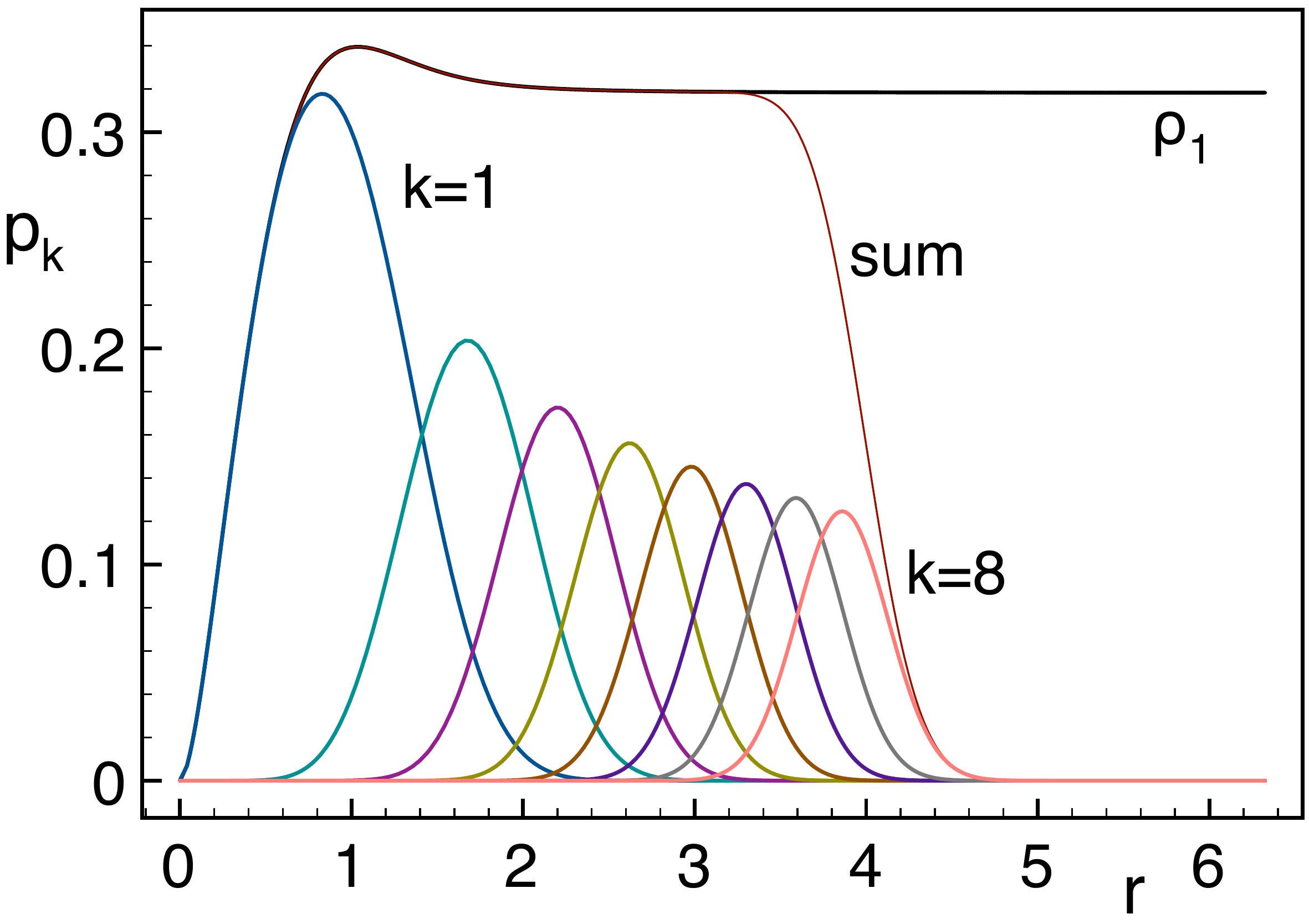}\hfill%
  \includegraphics[width=42mm]{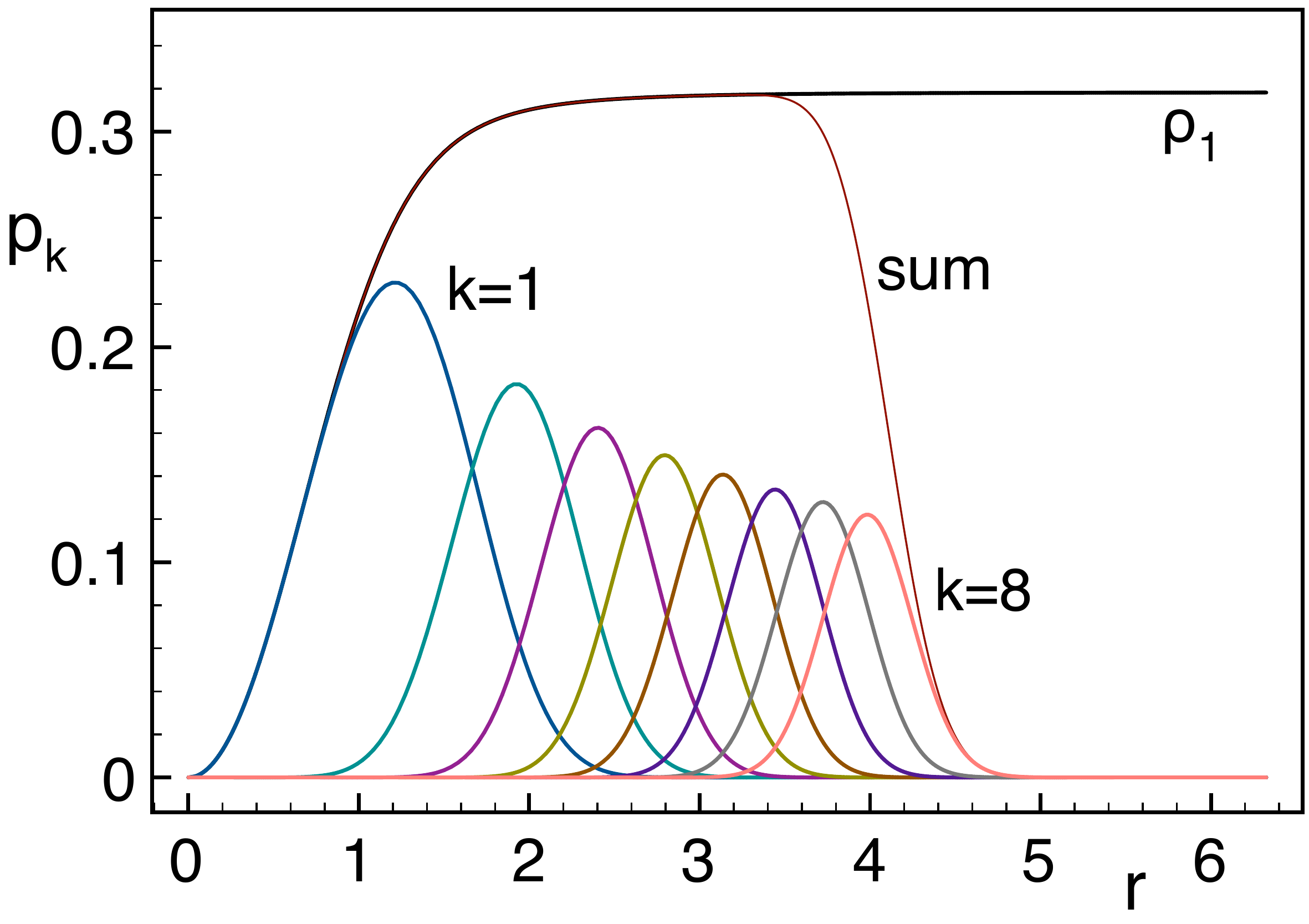}
  \caption{Quenched spectral density Eq.~\eqref{rhoQS} and
    distributions of the first eight eigenvalues Eq.~\eqref{pk}, as
    well as their sum, all in the large-$\alpha$ limit, for $\nu=0$
    (left) and $\nu=1$ (right).}
  \label{psum}
\end{figure}

Denoting each factor in Eq.~\eqref{E0} by $1-\lambda_\ell$,
expressions for the $E_k(r)$ easily follow in terms of the
$\lambda_\ell$ \cite{Mehta}.  The radially ordered eigenvalue
distributions are then obtained from the $E_k(r)$ via
Eq.~\eqref{Eprel}, leading to
\begin{equation}
  p_k(r) = -\frac1{\pi r}\frac{\partial}{\partial r} 
  \sum_{n=0}^{k-1} \frac{E_n(r)}{n!}.
\label{pk}
\end{equation}
Figure~\ref{psum} shows that the individual eigenvalue distributions
$p_k(r)$ nicely add up to the density Eq.~\eqref{rhoQS}.

\textbf{Comparison with lattice data.}  We now come to the comparison
of our analytical results to quenched lattice QCD data.  For details
of the simulation we refer to Ref.~\cite{Bloch:2006cd}.  The gauge
fields were generated in the quenched approximation on a $4^4$ lattice
at $\beta=5.1$ (see \cite{Bloch:2006cd} for an explanation of these
choices).  The Dirac operator introduced in Ref.~\cite{Bloch:2006cd}
is a generalization of the overlap Dirac operator \cite{overlap} to
$\mu\ne0$.  It satisfies a Ginsparg-Wilson relation
\cite{Ginsparg:1981bj} and has exact zero modes at finite lattice
spacing.  We can therefore test our predictions in different sectors
of topological charge $\nu$.  In Ref.~\cite{Bloch:2006cd} complete
spectra of the generalized overlap operator were computed for several
values of $\mu$ and large numbers of configurations, and these data
are used in the comparisons to the RMT results below.  We also used
the fit parameters $\Sigma$ and $\fpi$ from Ref.~\cite{Bloch:2006cd}
to determine $\alpha$ and $\xi$, i.e., no additional fits were
performed.

\begin{figure}[b]
  \centering
  \includegraphics[width=42mm]{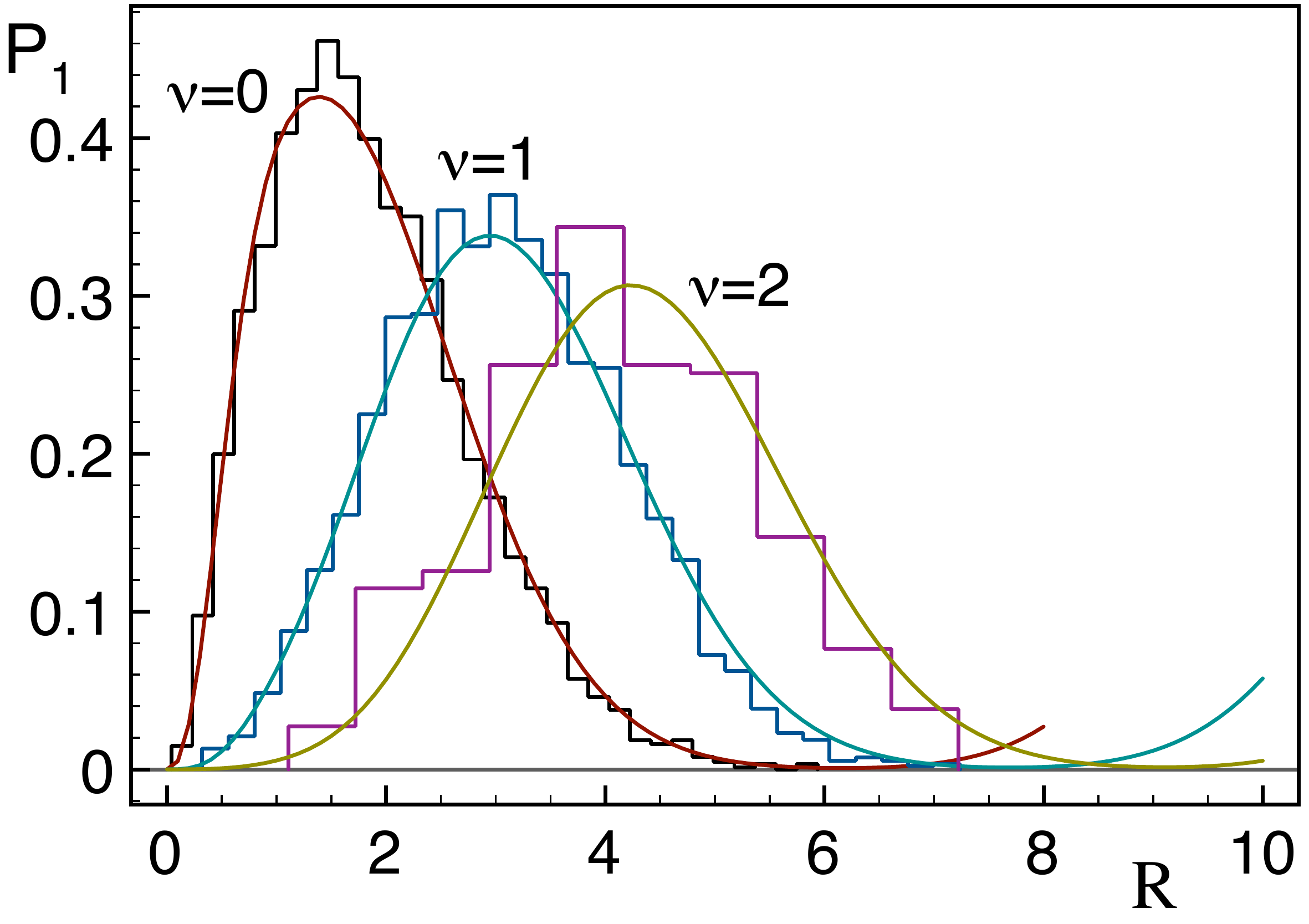}\hfill%
  \includegraphics[width=42mm]{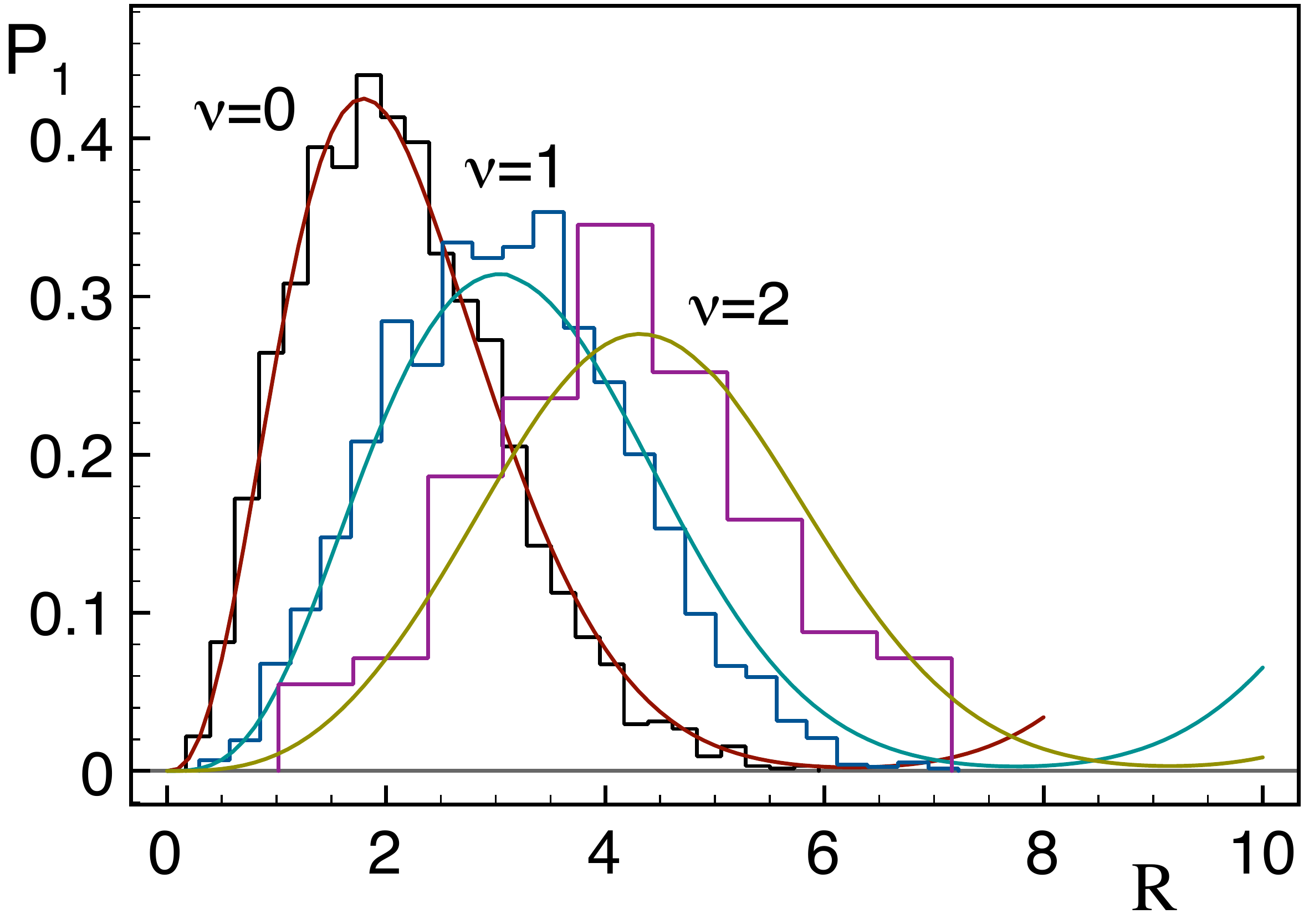}
  \caption{Integrated distribution $P_1(R)=\int_0^\pi
    d\theta\,R\,p_1(R,\theta)$ of the first eigenvalue for $\nu=0,1,2$
    as a function of the radius $R$ for $\mu=0.1$ (left) and $\mu=0.2$
    (right).  The solid lines are the RMT results from
    Eq.~\eqref{p1exp}, the histograms are the quenched lattice data of
    Ref.~\cite{Bloch:2006cd}.  The bending-up of the RMT curves for
    large $R$ is an artifact of using only the first three terms in
    the expansion \eqref{p1exp}, see text.}
  \label{datamu0.1-2}
\end{figure}

\begin{figure}[t]
  \centering
  \includegraphics[width=42mm]{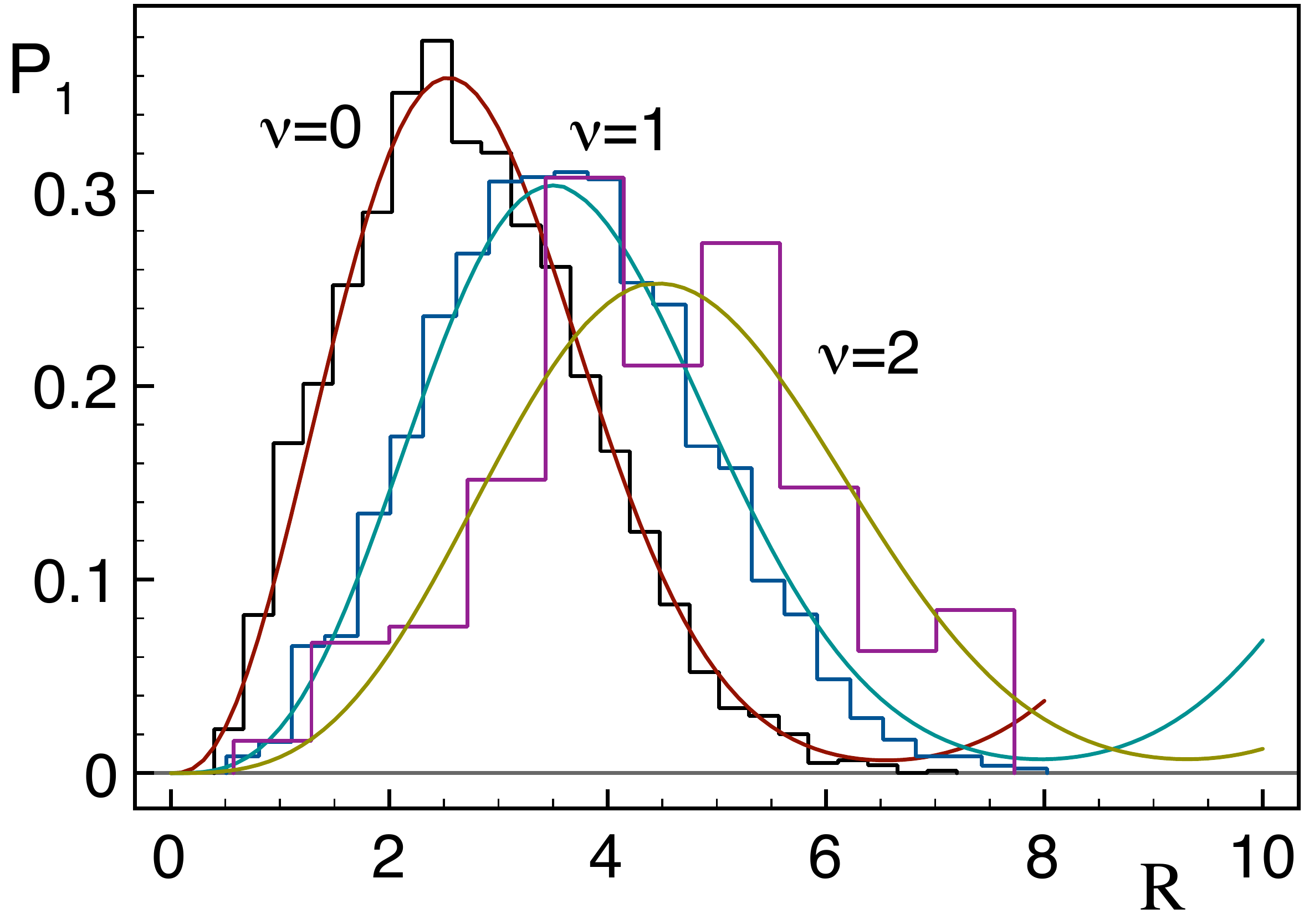}\hfill%
  \includegraphics[width=42mm]{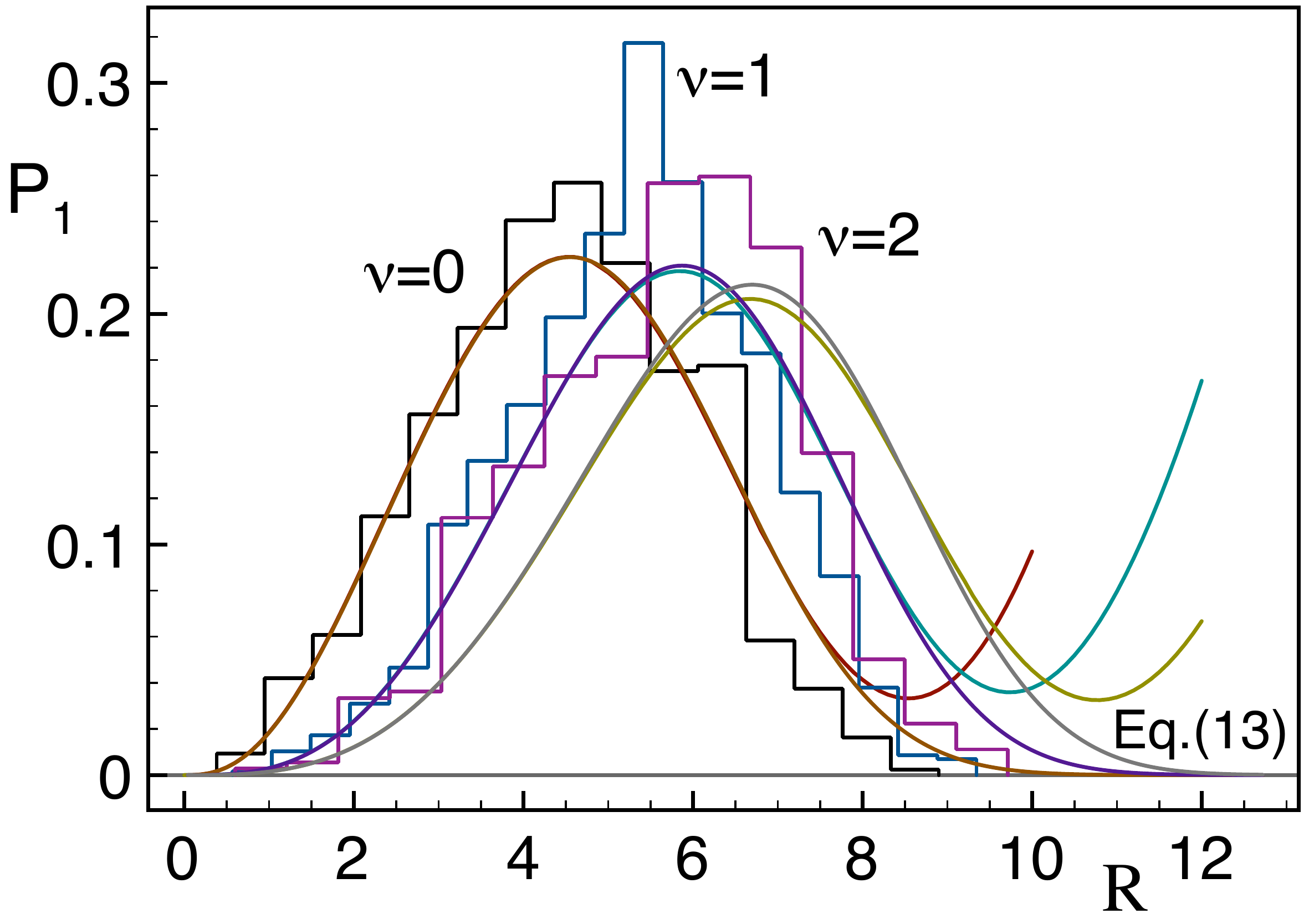}
  \caption{Same as Fig.~\ref{datamu0.1-2}, but for $\mu=0.3$ (left)
    and $\mu=1.0$ (right).  For $\mu=1.0$ we also show the exact RMT
    result in the large-$\alpha$ limit from Eq.~\eqref{pk}.}
  \label{datamu0.3+1}
\end{figure}

For the contours $\partial J$ we again choose semi-circles, for all
values of $\alpha$.  Since we prefer to show 2D plots we have
integrated over the phase of the complex number $\xi=Re^{i\theta}$ and
display only the radial dependence.  Results for $\nu=0,1,2$ are shown
in Fig.~\ref{datamu0.1-2} for $\mu=0.1$ and $\mu=0.2$, corresponding
to $\alpha=0.174$ and $\alpha=0.615$, and in Fig.~\ref{datamu0.3+1}
for $\mu=0.3$ and $\mu=1.0$, corresponding to $\alpha=1.42$ and
$\alpha=4.51$.  (The lattice spacing $a$ has been set to unity.)
For all values of $\mu$ we compare the data to the expansion
Eq.~\eqref{p1exp}, in which only the first three terms were used.

For $\mu=1.0$ the data were found to be approximately rotationally
invariant, and we also compare them to the exact result in the
large-$\alpha$ limit from Eqs.~\eqref{E0} and \eqref{pk}.  (Because of
the rotational invariance, only the ratio $\Sigma/\fpi$ could be
determined for $\mu=1.0$ in Ref.~\cite{Bloch:2006cd}, see
Eq.~\eqref{rhoQS}.  In this case the value of $\alpha$ used in
Eq.~\eqref{p1exp} is an extrapolation, assuming that $\Sigma$ is
independent of $\mu$.)  The agreement between data and analytical
curves is excellent except for $\nu=1,2$ at $\mu=1.0$ (see
Fig.~\ref{datamu0.3+1}).  In these two cases we have left the range of
validity of RMT.

We emphasize that while the rise of the distributions from zero was in
principle already tested in Ref.~\cite{Bloch:2006cd} through the
density (see Fig.~\ref{3d} or \ref{psum}), their decrease represents a
new, parameter-free test. Note also that because of the integration
over the phase, the signal is much better than in
Ref.~\cite{Bloch:2006cd}.  This allows us, for the first time, to
successfully test the RMT predictions for $\nu=2$.

Figures~\ref{datamu0.1-2} and \ref{datamu0.3+1} also show the effect
of truncating the Fredholm expansion \eqref{p1exp}: The analytical
curves bend up for large $R$ after (almost) touching zero.
Higher-order terms in the expansion \eqref{p1exp} only affect the tail
of the distributions.  They will ``repair'' the bending-up and ensure
that the tails remain zero, just as the data. The same effect was
observed earlier for real eigenvalue distributions
\cite{Akemann:2003tv}.  This feature of our approximation can be seen
most clearly when comparing to the exact result in the large-$\alpha$
limit, see Fig.~\ref{datamu0.3+1} (right), in which we can observe how
the expansion converges in the case of large $\alpha$.\\[-3mm]

\textbf{Conclusions.}  We have shown that the distributions of
individual complex eigenvalues from non-Hermitian RMT agree very well
with the corresponding distributions of the complex eigenvalues of the
quenched QCD Dirac operator closest to the origin in three different
topological sectors.  As in the Hermitian case, these distributions
are much easier to compare with than the density, in which a plateau
may not be observable due to appreciable finite-volume corrections.
Our analytical results are also relevant for other non-Hermitian
systems in the chiral unitary symmetry class.  In the future, it would
be interesting to compute (and apply) similar results for the
orthogonal and
symplectic symmetry classes.\\[-3mm]

\textbf{Acknowledgments.} We thank K. Splittorff and
J.J.M. Verbaarschot for useful discussions. 
This work was supported by EPSRC grant EP/D031613/1 (GA \& LS), by EU
network ENRAGE MRTN-CT-2004-005616 (GA), and by DFG grant FOR 465 (JB
\& TW).

\bibliography{ABSWv2}

\end{document}